\documentclass[letterpaper]{article}

\usepackage[T1]{fontenc}

\usepackage{geometry}
\usepackage{setspace}

\usepackage[articletitle=true]{achemso}

\usepackage[colorlinks=true,
linkcolor=blue,
citecolor=blue,
urlcolor=blue]{hyperref}
\usepackage{graphicx}
\usepackage{float}
\newfloat{scheme}{htbp}{los}
\floatname{scheme}{Scheme}
\floatname{chart}{Chart}
\newfloat{graph}{htbp}{loh}
\usepackage{amsmath,amssymb,amsfonts}%
\usepackage{mathrsfs}%
\usepackage[title]{appendix}%
\usepackage{xcolor}%
\usepackage{textcomp}%
\usepackage{booktabs}%
\usepackage{listings}%
\usepackage[mathlines]{lineno}
\usepackage{braket} 
\usepackage{soul} 
\usepackage{epstopdf}
\usepackage{etoolbox}
\usepackage{chemformula} 
\usepackage[version = 4]{mhchem} 

\usepackage{authblk}
\author[1,2]{Jiaru Zhou}
\author[1,2]{Wenze Lan}
\author[1]{Hao Li}
\author[1,2]{Yu Hua}
\author[1,2]{Peng Fu}
\author[1,2]{Geng Li}
\author[3]{Xiaofeng Fan}
\author[1,2]{Changzhi Gu*}
\author[1.2]{Baoli Liu*}

\affil[1]{Beijing National Laboratory for Condensed Matter Physics, Institute of Physics, Chinese Academy of Sciences, Beijing, 100190, P. R. China}
\affil[2]{School of Physical Sciences, CAS Key Laboratory of Vacuum Physics, University of Chinese Academy of Sciences, Beijing, 100190, P. R. China}
\affil[3]{College of Materials, Jilin University, No.2699 Qianjin Street, Changchun, 130012, P. R. China}

\title{Re-evaluation of bottleneck effect via a coupled  monolayer WS$_2$/photonic crystal heterostructure}
\date{*blliu@iphy.ac.cn; czgu@iphy.ac.cn}

\begin{document}
\maketitle

\begin{abstract}
Exciton-polariton condensates is an important type of Bose–Einstein condensate whose realization requires efficient relaxation of polaritons to the band-energy minima. However, this process is often obstructed by bottleneck effect near the anticrossing region of polariton dispersion. Although the exciton-polariton bottleneck effect has been extensively observed in various polariton system, but there is no a unified views of physical origin. Here, we construct an exciton–trion–photon coupling system in monolayer WS$_2$/photonic-crystal slab heterostructures. Momentum-resolved photoluminescence reveals the anticrossing polariton dispersions for the exciton resonance with a $\sim$57~meV Rabi splitting and there is no characteristic anticrossing for trion resonance with a $\sim$5~meV splitting at $\sim$12~K. Enhanced polariton emission is observed around the trion–polariton crossing with elevating temperature. We attributes this exotic phenomenon to bottleneck effect and indicating that small Rabi splitting is the unified origin of bottleneck effect in polariton systems.
\end{abstract}

\section*{Keywords}

bottleneck effect, exciton polariton, trion, photonic crystal



\section{Introduction}

The interaction between light and matter is a central topic in modern optics. It occurs not only in atomic and solid-state systems but also shows distinct features in nanophotonics. When the rate of energy exchange between photons and matter excitations exceeds their decoherence rates, the system enters into the strong coupling regime\cite{Khitrova2006}. Under this condition, light and matter couple to form polaritons, that is, bose-type quasi-particles with both light and matter properties\cite{DengHui2010}. As a hybrid state of photons and excitons, exciton-polaritons inherit their advantages, including very low effective mass, strong nonlinear effects, high-speed  propagation, and high sensitivity to regulation of external fields (such as electric and magnetic fields). These advantages make them an important link between condensed matter systems and photonic systems, with importance in both fundamental physics and optoelectronic devices\cite{Sanvitto2016}. In terms of fundamental physics, exciton-polaritons are promising candidates for studying collective excitation at room temperature\cite{XiongQihua2021}, such as unequilibrium Bose-Einstein condensation\cite{Kasprzak2006}, superfluidity\cite{Amo2009, Lerario2017}, quantum vortices\cite{Lagoudakis2008}, and low-threshold polariton lasing\cite{Cohen2010}. They also show potential in quantum simulation, computing, and information processing, as well as in the development of qubits\cite{Ghosh2020}.
\\

The exciton-polariton condensates has been realized  in GaAs and CdTe materials at cryogenic temperature of $\sim$10~K \cite{Kasprzak2006,Pfeiffer2007,Deng2002}, and in other materials such as GaN \cite{Christopoulos2007,Baumberg2008}, ZnO\cite{Guillet2011}, transition metal dichalcogenides (TMDCs)\cite{Sanvitto2022,Wu2023}, perovskites\cite{Xiong2020}, and organics\cite{Rainer2014} even at higher temperature. For  exciton-cavity system, the fundamental requirement of realizing exciton-polariton condensates is that the exciton-polaritons must relax effectively and rapidly to the energy minima located in polariton energy bands. However, the high-energy exciton polaritons sometimes can not relax to the energy minima and spread out the lower polariton branch (LP) \cite{Yamamoto2002,Dang2005,OE2011,Laitz2023,Coles2013}. Those kinds of phenomena were basically attributed to the bottleneck effect\cite{Pau1995,Tassone1997}.   Although the physical mechanisms of the carrier-carrier scattering \cite{Tartakovskii2000} and LO-phonon-mediated relaxation \cite{Laitz2023} were proposed to release the bottleneck, there is no unified understanding about the origin of the bottleneck effect.  
\\

In classical definition of bottleneck effect, the steep polariton dispersion resulting from Rabi splitting around the momentum of exciton-photon resonance and the decrease of the relaxation rate in passing from the excitonlike to photonlike polaritons were the  main origin of this effect \cite{Yamamoto2014,Tassone1997}. However,  the steepness of polariton dispersions strongly depends on the magnitude of exciton-photon coupling  which can be characterized by Rabi splitting. Furthermore, in the pursuit of  exciton-polariton condensates, the small Rabi splitting is always ignored because it is hard to identify the generation of exciton polaritons due to the indistinguishable anticrossing characteristics in polariton dispersions
\cite{Barnes2015}.  Therefore, does the bottleneck effect occur for any magnitude of Rabi splitting? What is the role of the small Rabi splitting in the bottleneck effect? Up to now, both issues still remain to be elucidated clearly in order to understand deeply the origin of the bottleneck effect in polariton system.

\section{Results and discussion}
\begin{figure}[t]
	\centering
	\includegraphics[width=0.6\textwidth]{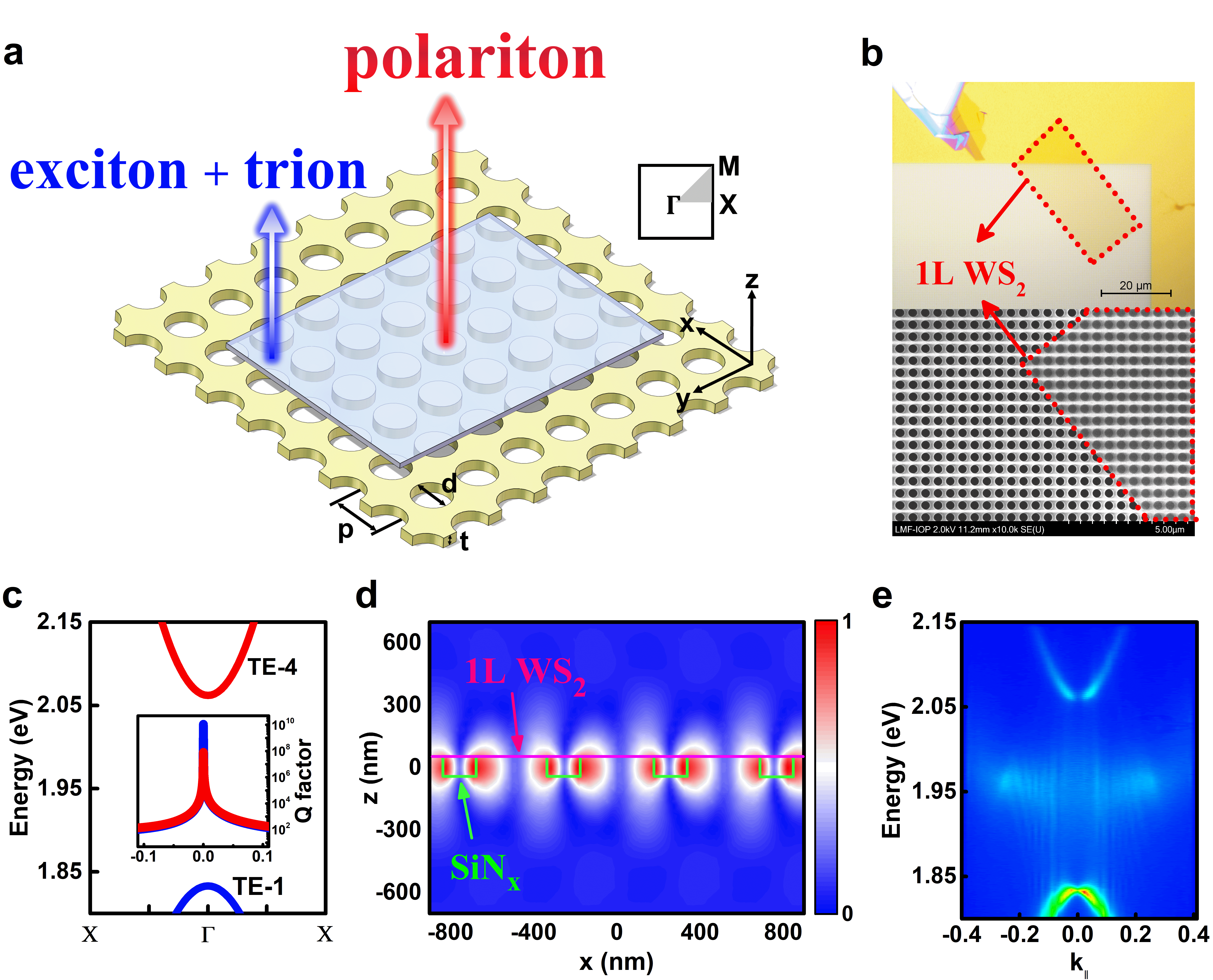}\\
	\caption{\textbf{The coupling system.}(a) Schematic illustration of the two-dimensional PhC slab with a layer of monolayer (ML) WS$_2$  on the top. (b) Optical microscope image of the structure (top) and a top-view scanning electron microscope (SEM) image of the device (bottom). The dotted red line indicates the contour of ML WS$_2$. (c) Simulated TE-like photonic band structure of the TE$_1$ and TE$_4$ modes of the bare PhC (without the WS$_2$ monolayer). The inset shows the corresponding quality factors (Q factors). (d) Calculated electric field intensity of the TE$_4$ mode (normalized to its maximum value) at an in-plane momentum of $k_{\parallel}=0.10$ in x-z plane. The green and pink lines mark the outline of the PhC and the WS$_2$ monolayer, respectively. (e) Momentum-resolved photoluminescence spectra of the bare PhC, exhibiting a sharp, dispersive cavity mode, in good agreement with the simulated results in (c).}
	\label{Figure 1}
\end{figure}

In this work, we create a exciton-photon coupling system which consists of a monolayer (ML)  tungsten disulfide (WS$_2$) and a photonic crystal (PhC) slab  as shown in Figure~\ref{Figure 1}a. The ML WS$_2$ provides two types of excitons: neutral excitons and charged excitons (trion), which exhibit different coupling strengths with photons\cite{Emmanuele2020}. Hereinafter, neutral  exciton and charged exciton are represented by exciton and trion, respectively. The PhC is patterned into a square lattice of cylindrical air holes in a silicon nitride (SiN$_\mathrm{x}$) membrane with a thickness of about 100 nm. ML WS$_2$ is transferred onto the surface of PhC through dry method. Figure~\ref{Figure 1}b displays an optical micrograph of the device together with a top-view scanning electron microscopy (SEM) image. The WS$_2$ monolayer position is highlighted by the dotted red contour. The bare PhC slab supports optical singularities known as symmetry-protected bound states in the continuum (BIC) at the Brillouin-zone center ($\Gamma$ point), which arise from the in-plane inherent C$_2$ rotational symmetry\cite{Hsu2016}. The simulated photonic band structures of the PhC slab show TE-like modes which guarantee the maximal coupling between exciton and cavity photon as shown in Figure~\ref{Figure 1}c\cite{LanWZ}. Notably, the simulated quality factors show that the TE-1 and TE-4 modes exhibit ultra-high quality factors at the $\Gamma$ point, revealing the existence of $\Gamma$-BICs\cite{LanWZ}. The normalized in-plane electric-field intensity profile of the TE-4 mode at an in-plane momentum $k_{\parallel}=0.10$ is shown in Figure~\ref{Figure 1}d, illustrating strong field confinement within the air hole of PhC. The property of cavity array in PhC ensures the simultaneous observation of exciton, trion and polariton emission for one spectrum as shown in Figure~\ref{Figure 1}a, which is the unique advantage of this coupling system. The momentum-resolved photoluminescence  (PL) spectra of bare PhC were measured along the $\Gamma$-X direction at room temperature utilizing the 4f spectroscopy technique\cite{LanWZ} as presented in Figure~\ref{Figure 1}e. The measured photonic dispersions are in good agreement with the simulated results as shown in Figure~\ref{Figure 1}c. Here, $k_{\parallel}$ is a dimensionless in-plane wave vector (momentum) and equivalent to $k/k_{0}$, where $k=k_{0}\sin\theta$, $k_{0}=2\pi/\lambda$, $\theta$ is the angle with respect to sample normal.
\\

\begin{figure}[t]
	\centering
	\includegraphics[width=0.95\textwidth]{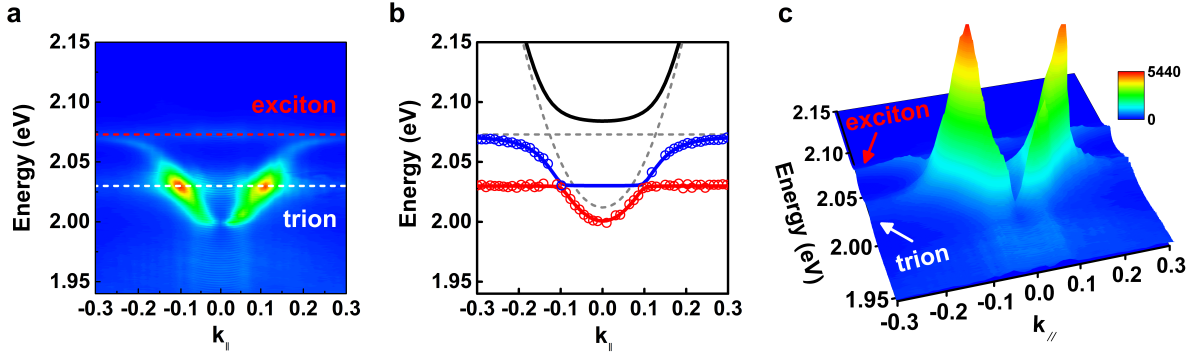}\\
	\caption{\textbf{Temperature-dependent polariton dispersions in the WS$_2$-PhC system.} (a) Momentum-resolved PL spectra of the WS$_2$-PhC integrated device measured at $\sim$12~K. (b) Polariton energies $E_{\mathrm{MP}}$ (blue circles) and $E_{\mathrm{LP}}$ (red circles) extracted from the spectra in (a). The fitted polariton dispersions obtained from a coupled oscillator model are shown as solid curves: upper, middle and lower polariton branches (UP, MP and LP) at 12~K, plotted in black, blue and red, respectively, giving a Rabi splitting $2\hbar\Omega_{\mathrm{cav,X}}=$ 57~meV, $2\hbar\Omega_{\mathrm{cav,T}}=$ 5~meV at $\sim$12~K. Gray dashed lines indicate the uncoupled cavity, exciton and trion energies, $E_{\mathrm{cav}}$, $E_{\mathrm{T}}$ and $E_{\mathrm{exc}}$. (c) 3D representations of the momentum-resolved PL spectra shown in (a) which shows a pronounced enhancement near the crossing between the lower polariton branch and trion resonance at $\sim$12~K.}
	\label{Figure 2}
\end{figure}

We characterize the ML WS$_2$-PhC device with various temperatures via momentum-resolved PL spectroscopy along the $\Gamma$-X direction. Figure~\ref{Figure 2}a presents the momentum-resolved PL spectra of the ML WS$_2$-PhC device at low temperature of $\sim$12~K. We observed simultaneously the emission of polariton, dispersionless exciton and trion due to a unique advantage of our coupling system. The peak energies of exciton and trion are $\sim$ 2.07~eV and  $\sim$ 2.03~eV, respectively  (See Figure S3 for extraction method in supplementry materials), which is in agreement with the results of the literature\cite{Tobias2015}. The anticrossing dispersions of polaritons obviously can be observed for exciton-photon resonance while there is no clear anticrossing for trion-photon resonance due to weak coupling between trion and cavity photon\cite{Tartakovskii2017}. The Rabi splitting of trion-photon coupling is buried by the larger half linewidth of trion peak which is around $\gamma_{\mathrm{T}}\approx$ 13~meV\cite{Barnes2015}. The three-oscillator models of exciton-trion-photon was exploited to fit the experimental dispersions of PL spectra with following Hamiltonian:
\begin{equation}
	H=
	\begin{pmatrix}
		E_{\mathrm{cav}}(k_{\parallel})-i\gamma_{\mathrm{cav}}  &g_{\mathrm{exc}}                        &g_{\mathrm{T}}\\
		g_{\mathrm{exc}}                           &E_{\mathrm{exc}}-i\gamma_{\mathrm{exc}}  &0\\
		g_{\mathrm{T}}                           &0                              &E_{\mathrm{T}}-i\gamma_{\mathrm{T}}\\
	\end{pmatrix}	
	\label{eq:3}\\
\end{equation}
where $g_{\mathrm{exc}}$, $g_{\mathrm{T}}$ are the coupling strengths of the photon-exciton and photon-trion, respectively. The fitting details are described in the \nameref{method} section. The fitting curves are plotted in Figure~\ref{Figure 2}b, showing excellent agreement with experimental data with Rabi splitting of $\sim$57~meV for exciton-photon coupling and $\sim$5~meV for trion-photon coupling, respectively. The small Rabi splitting of $\sim$5~meV is comparable with previous report\cite{Tartakovskii2017}. 
\\

An exotic result in this spectra is that the stronger light emission can be observed around $k_{\parallel}=\pm 0.10$ which is the cross point of trion and a branch of polarion in k-space as shown in Figure~\ref{Figure 2}a and more clearly in Figure~\ref{Figure 2}c. This phenomenon is never reported before. In general, the stronger PL emission mainly comes from the Purcell effect\cite{Purcell1946} and/or the accumulation of  "emitters" . In our case, the momentum of cavity mode with the trion energy of $\sim$2.03~eV is around $k_{\parallel}=\pm 0.068$ which is far from the momentum of PL enhancement $k_{\parallel}=\pm 0.10$ as presented in Figure~\ref{Figure 2}b. Therefore, the Purcell effect of trion emission can be excluded for this light emission enhancement.  In this ML WS$_2$-PhC device, the accumulation of trions also can be ruled out. Consequently, we identify that the polariton's accumlation is the main origin of stronger light emission. This means that the relaxation process of the high-energy polaritons was blocked intensely around the cross point of $k_{\parallel}=\pm 0.10$ and the trion energy of $\sim$2.03~eV. And then, the polaritons can not relax efficiently to minima of polariton dispersion. We attribute this huge suppression of the polaritons relaxation to the well-known bottleneck effect in polariton system \cite{Yamamoto2014,Tassone1997,Menon2022}. 
\\

\begin{figure}[H]
	\centering
	\includegraphics[width=0.6\textwidth]{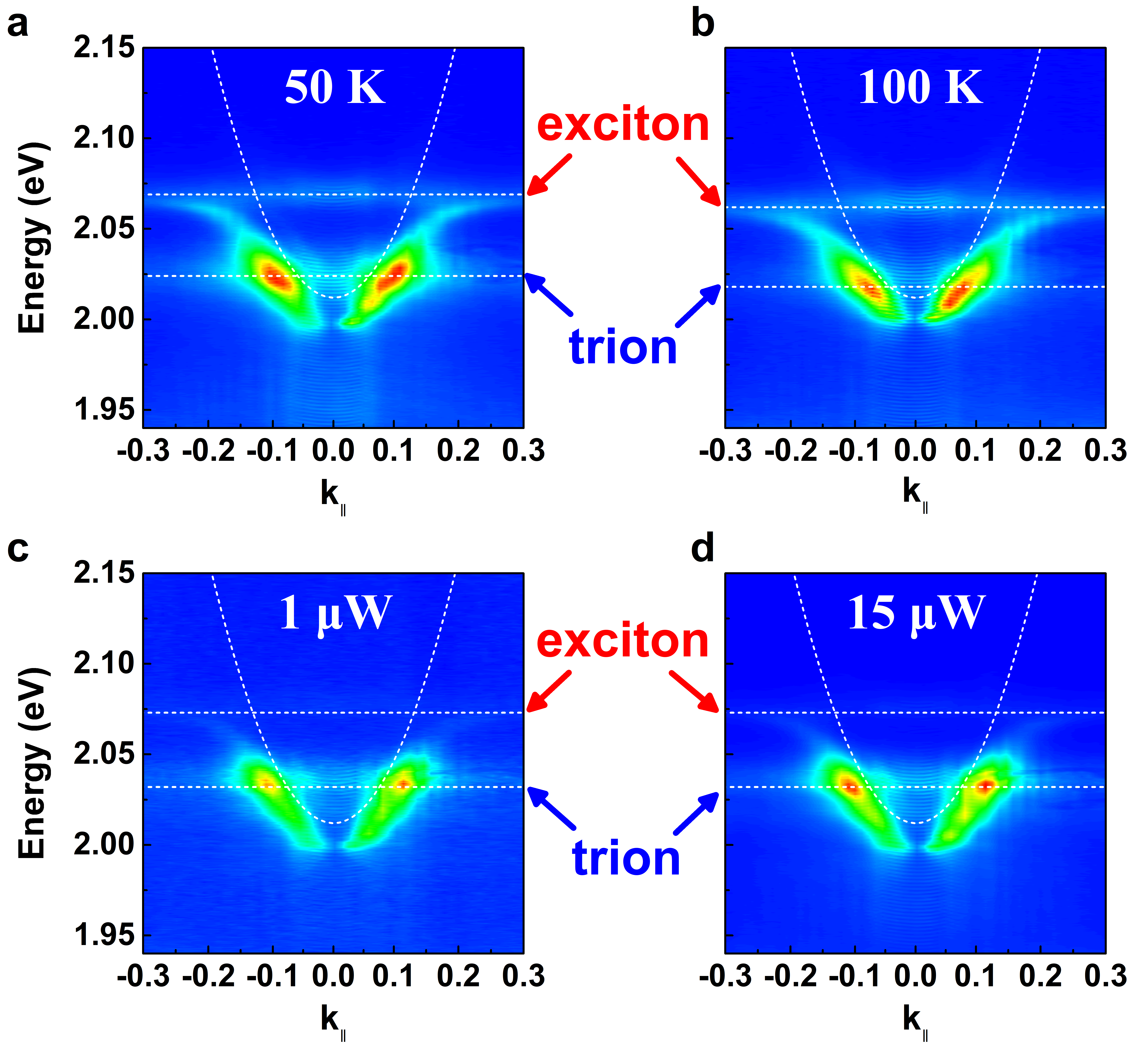}\\
	\caption{\textbf{Momentum-resolved polariton emission under varying temperature and excitation power.}(a, b) Momentum-resolved PL spectra of the WS$_2$-PhC integrated device measured at temperatures of $\sim$50~K and $\sim$100~K, respectively, illustrating the temperature-dependent evolution of the polariton emission. (c, d) Momentum-resolved PL spectra recorded under different excitation powers of $\sim$1~\textmu W and $\sim$15~\textmu W, respectively.}
	\label{Figure 3}
\end{figure}

In order to confirm that this bottleneck effect results from the weak coupling of trion and cavity photon, we perform the measurements of the momentum-resolved PL spectroscopy with elevating temperature. Figure~\ref{Figure 3}a and b show the temperature-dependent spectra of ML WS$_2$-PhC device at temperature of $\sim$50~K and $\sim$100~K, respectively. It is clear that the momentum position of stronger polariton emission in k-space still be seated at the cross point of trion and polariton. We can confirm the bottleneck effect results from the weak coupling between trion and cavity photon although the anticrossing or Rabi splitting is indistinguishable in polariton spectra.
\\

Let us recall the original definition of bottleneck effect in polariton system\cite{Pau1995,Tassone1997}, the steepened dispersions of polaritons due to coupling between exciton and cavity photon is the dominant mechanism of the bottleneck effect. Therefore, we check what happens around the anticrossing regimes of neutral exciton interacting strongly with cavity photon. As shown in Figure~\ref{Figure 2}a, Figure~\ref{Figure 3}a$\And$b, the intensity of polariton emission in anticrossing region is weaker comparing to that of low polariton branch. This implies that the bottleneck effect is invalid although the anticrossing dispersions can be observed distinctly. One possibility of releasing bottleneck effect is the enhanced  polariton-polariton scattering with higher polariton density\cite{Tartakovskii2000} under excitation power of $\sim$10~\textmu W in our experiment. We perform the measurements of the power-dependent spectra of  ML WS$_2$-PhC device  as shown in Figure~\ref{Figure 3}c-d  with excitation power varying from $\sim$1~\textmu W to $\sim$15~\textmu W. Again the polariton emission is weaker around the anticrossing regions although the anticrossing dispersions induced by strong coupling of neutral exciton and cavity photon are clear for all excitation powers.  This means that there does not exist the bottleneck effect around those anticrossing regions. Based on all above mentioned analysis, we can make a solid conclusion that the small Rabi splitting is the true origin of bottleneck effect in our  ML WS$_2$-PhC device. This conclusion is consistent to recent experimental observation in which the small Rabi splitting indeed induced the polariton accumulation\cite{Menon2022}. The Rabi splitting less than $\sim$ 5~meV can introduce the bottleneck effect, while there is no bottleneck effect when the Rabi splitting is larger than $\sim$ 57~meV. The effect of intermediate coupling strength (Rabi splitting in the range of 5$\sim$57~meV) on the bottleneck effect needs to be further clarified experimentally in the future.
\\

\begin{figure}[t]
	\centering
	\includegraphics[width=1\textwidth]{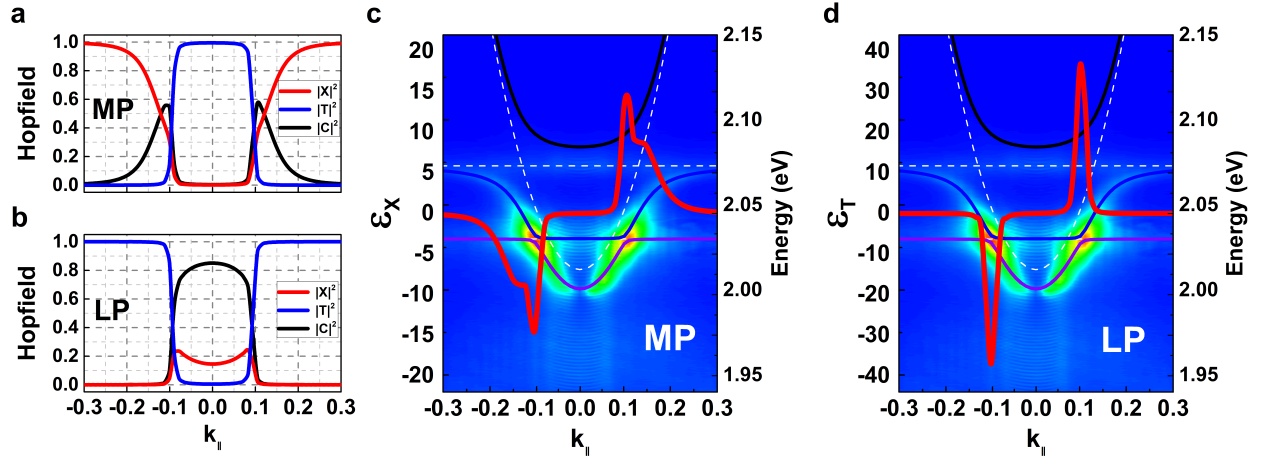}\\
	\caption{\textbf{Hopfield composition and its momentum-dependent variation in the exciton–trion–photon polariton system}(a, b) Hopfield coefficients of the middle polariton (MP) and lower polariton (LP) branches at $\sim$12~K, respectively. The exciton ($|\mathrm{X}|^2$, red) and trion ($|\mathrm{T}|^2$, blue) and photon ($|\mathrm{C}|^2$, black) fractions are obtained from the three-mode-coupled oscillator model, illustrating the composition evolution of the polariton states as a function of in-plane momentum  $k_{\parallel}$. (c) the rate of change of excitonic Hopfield coefficients  $\varepsilon_\mathrm{X}$ (red curve) in the MP branch, extracted from panel (a), plotted together with the fitted dispersions of UP, MP and LP branches. (d) the rate of change of trionic Hopfield coefficients  $\varepsilon_\mathrm{T}$ (red curve) in the LP branch, extracted from panel (b), plotted together with the fitted dispersions of UP, MP and LP branches. }
	\label{Figure 4}
\end{figure}

Finally, we discuss the physical mechanism of the bottleneck effect in this exciton-trion-photon coupling system. That the polaritons relax to lower energy states is the scattering process which is principally determined by the matter properties of polaritons. Here matter refers to  exciton and trion. The relaxation efficiency of polaritons  predominately  depends on the difference of density of states (DOS) of exciton/trion components between final and initial polariton states for the case of polariton scattering. For the sake of simplicity, Hopfield coefficients\cite{Hopfield1958} of exciton and trion in polariton states can be used to take the place of the DOS of exciton and trion very well. Because the stronger polariton emission happens around the cross point of trion and polariton dispersion, Hopfield coefficients of excitonic, trionic and photonic fractions were calculated only for middle polariotn branch (MP)  and lower polariotn branch (LP) as presented in Figure~\ref{Figure 4}a and Figure~\ref{Figure 4}b, respectively. In our coupling system, the polaritons relax from  high-k to lower-k states in k-space, so far we just consider the case of Hopfield coefficient's decrease for excitonic and trionic fractions.
\\

For excitonic component in MP, Hopfield coefficient decreases gently from high-k to  $k_{\parallel}=\pm 0.12$ while this coefficient drops sharply from  $k_{\parallel}=\pm 0.12$ to zero as shown by red curve in Figure~\ref{Figure 4}a. Meanwhile, the trionic Hopfield coefficients decrease abruptly around  $k_{\parallel}=\pm 0.10$ for LP as presented by blue curve in Figure~\ref{Figure 4}b. In order to quantify the variation degree of  both excitonic and trionic Hopfield coefficients, we define $\varepsilon_\mathrm{X}=\frac{d|\mathrm{X}|^2}{dk_{\parallel}}$ and $\varepsilon_\mathrm{T}=\frac{d|\mathrm{T}|^2}{dk_{\parallel}}$ as the  rate of change of Hopfield coefficients as a function of  $k_{\parallel}$. Here, $|\mathrm{X}|^2$ and $|\mathrm{T}|^2$ are excitonic and trionic Hopfield coefficients, respectively. Figure~\ref{Figure 4}c presents the rate of change of excitonic Hopfield coefficients  $\varepsilon_\mathrm{X}$.  It is clear that $|\varepsilon_\mathrm{X}|$ varies smoothly and is less than $\sim$10 when $k_{\parallel}$ varies from high-k to $k_{\parallel}=\pm 0.12$. This range of $k_{\parallel}$ values exactly corresponds to the anticrossing regions of exciton-photon strong coupling and  no bottleneck effect happens as shown in Figure~\ref{Figure 4}c. When $|k_{\parallel}|$ is less than $\sim$0.12, $|\varepsilon_\mathrm{X}|$ changes dramatically and reaches the maximum of $\sim$ 15 at  $|k_{\parallel}|=0.10$. The stronger polariton emission just right emerges around  $|k_{\parallel}|=0.10$ as presented Figure~\ref{Figure 4}c and the bottleneck effect plays a key role. Now, we turn to explore the properties of the rate of trionic Hopfield coefficient change  $|\varepsilon_\mathrm{T}|$ in LP dispersion. Figure~\ref{Figure 4}d shows the curve of $\varepsilon_\mathrm{T}$ as a function of  $k_{\parallel}$. In contrast to  $|\varepsilon_\mathrm{X}|$,  $|\varepsilon_\mathrm{T}|$ only changes abruptly around  $|k_{\parallel}|=0.10$ and gets the maximum of $\sim$ 40 at $|k_{\parallel}|=0.10$, around which the bottleneck effect takes effect again. So far, the sharp change of excitonic and trionic Hopfield coefficients is the main physical mechanism of the bottleneck effect. The rates of  change of excitonic and trionic Hopfield coefficients $|\varepsilon_\mathrm{X}|$ and  $|\varepsilon_\mathrm{T}|$ can be utilized well to evaluate the bottleneck effect through the emergence of larger  absolute extrema with narrow linewidth in $|\varepsilon_\mathrm{X}|$ and  $|\varepsilon_\mathrm{T}|$ in our polariton system. 
\\

In fact, monolayer WS$_{2}$  provides an ideal and controllable  platform for investigating the relationship between the bottleneck effect and the exciton-photon coupling strength. This material hosts two distinct types of excitons that couple to photons with markedly different strengths and can be spectrally distinguished, allowing for a direct comparison of polariton dynamics under strong and weak coupling conditions within the same device. In contrast, in other material systems where the bottleneck effect is commonly observed - such as ZnO, perovskites and others - the target excitons are typically accompanied by additional excitonic species that couple weakly to photons and often cannot be distinguished in the spectrum. This fundamental limitation explains why a unified origin of the bottleneck effect has remained elusive in previous studies.
\\

In summary, the stronger polariton emission was observed in exciton-trion-photon coupling system in a ML WS$_2$-PhC device.  We demonstrate that the small Rabi splitting/weak coupling strength is the unified origin of the bottleneck effect and this criteria is suitable to any polariton system. This finding  provides a novel scenario to overcome the bottleneck effect - creating pure exciton collectives, which couple to photons with same coupling strength and the larger Rabi splitting in pursuing exciton polariton condensates. Furthermore, this finding also offers significant potential for polariton applications, especially electrically controllable  exciton polariton condensates devices.

\section*{Methods}\label{method}
\subsection{Numerical simulation} A 3D finite element method  was used to perform the simulations in this work. The three-dimensional model was created with the periodic condition imposed in the x and y directions, and two perfect-matching layers (PML) in z direction. The band structures, eigenmodes and Q factors of the PhC slab were calculated using the "eigenfrequency" module in the frequency domain. The refractive index of SiN$_{\mathrm{x}}$ is set to $\sim$2.2.

\subsection{Sample fabrication} Photonic crystal slab shown in Figure~\ref{Figure 1}a was fabricated on commercial SiN$_{\mathrm{x}}$ membranes (NX5100c, Norcada) with the parameters: thickness t=100~nm, lattice period p = 504 nm, and diameter of air hole d = 372 nm. The membrane is based on silicon substrates with a central window of 1~mm $\times$ 1~mm, framed by a 5~mm$\times$5~mm silicon chip of 200~\textmu m thickness. The SiN$_{\mathrm{x}}$ layer was thoroughly etched to form the PhC modes. The SiN$_{\mathrm{x}}$ window was first spin-coated with a positive electron-beam resist (poly(methyl methacrylate) (PMMA) 495 A5, MicroChem). By spinning at 4000~r.p.m. for 60~s, a PMMA layer with a thickness of $\sim$ 200~nm was formed. The sample was then baked on a hotplate at 180~$^\circ\mathrm{C}$ for 60~s to remove residual solvent. 
A square lattice of cylindrical air holes was patterned by electron-beam lithography (EBL, Raith150). After exposure, the sample was developed in a mixture of methyl isobutyl ketone and isopropyl alcohol (MIBK:IPA = 1:3) for 40~s, followed by rinsing in IPA for 30~s. 
The periodically patterned PMMA layer served as an etching mask for the subsequent reactive ion etching (RIE) process. Anisotropic etching of the SiN$_{\mathrm{x}}$ layer was carried out using a CHF$_3$/O$_2$ gas mixture (50/5 sccm) at a chamber pressure of 55 mTorr and an RF power of 200 W. Finally, the remaining PMMA mask was removed by oxygen plasma etching (O$_2$, 50~sccm, 100~mTorr, 100~W).  Monolayer WS$_2$ flakes were exfoliated mechanically from bulk WS$_2$ materials purchased from 2D Semiconductors company and further transferred onto the surface of the SiN$_{\mathrm{x}}$ PhC using PDMS-assisted dry transfer technique.

\subsection{Optical measurements} Optical measurements were carried out using a custom-built momentum-resolved optical spectroscopy system based on a Fourier-space imaging configuration, integrated with a closed-cycle cryostat (Cryostation C2, Montana Instruments), providing temperature control from room temperature down to $\sim$12~K. The sample was mounted inside the cryostat on a piezo-controlled five-degree-of-freedom positioning stage and optically accessed through a 50$\times$ objectives (NA=0.4, Nikon), corresponding to an angular range of $\sim\pm 23.5^\circ$. The same objectives was used for both excitation and signal collection. A 532~nm continuous wave laser (MGL-III-532, CNI) served as the excitation source. In the momentum-resolved configuration, the back focal plane of the objectives was imaged onto the entrance slit of a spectrometer (iHR550, Horiba). The emitted signals were recorded by a two-dimensional charge-coupled device (CCD) array (Symphony, Horiba).

\subsection{Three-modes coupled harmonic oscillator model} To describe the formation of exciton-trion polaritons in the integrated device at lower temperatures, we employ a three-mode coupled oscillator model including the cavity photons, excitons and trions. In the basis of the cavity photon state $\ket{c}$, the exciton state $\ket{X}$ and the trion state $\ket{T}$, the corresponding eigenvalue equation reads:
\begin{equation}
	\begin{pmatrix} 
		E_{\mathrm{cav}}({\mathrm{k_{\parallel}}})-i\gamma_{\mathrm{cav}}  &g_{\mathrm{exc}}   &g_{\mathrm{T}}\\
		g_{\mathrm{exc}}     &E_{\mathrm{exc}}-i\gamma_{\mathrm{exc}}    &0\\
		g_{\mathrm{T}}     &0     &E_{\mathrm{T}}-i\gamma_{\mathrm{T}}
	\end{pmatrix}
	\begin{pmatrix}
		\mathrm{C}_n(k_{\parallel})\\
		\mathrm{X}_n(k_{\parallel})\\
		\mathrm{T}_n(k_{\parallel})\\
	\end{pmatrix}
	=E_n
	\begin{pmatrix}
		\mathrm{C}_n(k_{\parallel})\\
		\mathrm{X}_n(k_{\parallel})\\
		\mathrm{T}_n(k_{\parallel})\\
	\end{pmatrix}
	\label{eq:4}\\
\end{equation}
where $E_{\mathrm{cav}}(k_{\parallel})=E_0+\mathrm{a}k_{\parallel}^2$ denotes the bare cavity photon dispersion, $E_{\mathrm{exc}}$ and $E_{\mathrm{T}}$ are the bare exciton and trion energies, and $\gamma_{\mathrm{cav}}$, $\gamma_{\mathrm{exc}}$ and $\gamma_{\mathrm{T}}$ represent their corresponding half linewidths. The parameters $g_{\mathrm{exc}}$ and $g_{\mathrm{T}}$ describe the coupling strengths of cavity photon-exciton and cavity photon-trion resonances, respectively. $E_n$ are the eigenvalues corresponding to the UP, MP, LP branches. The eigenvector $(\mathrm{C}_{n}(k_{\parallel}), \mathrm{X}_{n}(k_{\parallel}), \mathrm{T}_{n}(k_{\parallel}))^\mathrm{T}$ describes the polariton state as a coherent superposition of photon, exciton and trion components, $\ket{n} = \mathrm{C}_{n}(k_{\parallel})\ket{c} + \mathrm{X}_{n}(k_{\parallel})\ket{X}+\mathrm{T}_{n}(k_{\parallel})\ket{T}$. Due to the presence of three coupled resonances, the eigenvalue problem does not admit a simple analytical solution. Therefore, the polariton dispersions were obtained by numerically diagonalizing the coupled-mode Hamiltonian at each in-plane momentum.

In our momentum-resolved PL measurements, the $E_{\mathrm{UP}}$ exhibits weak emission and cannot be extracted clearly. Therefore, the fitting procedure was performed using only the experimentally obtained $E_{\mathrm{MP}}$ and $E_{\mathrm{LP}}$. For a given set of coupling parameters ($g_{\mathrm{exc}}^0$, $g_{\mathrm{T}}^0$), the theoretical dispersions $E_n^{\mathrm{theory}}(k_{\parallel})$ were calculated and used to guide the identification of the corresponding spectral peaks in the experimental spectra. At each momentum, only the dominant emission peak located within a predefined energy window around the theoretical polariton position was selected as $E_n^{\mathrm{extract}}(k_{\parallel})$. 
The coupling strengths $g_{\mathrm{exc}}$, $g_{\mathrm{T}}$ were then determined through an iterative optimization procedure by minimizing the deviation between the experimentally extracted polariton energies and the calculated dispersions. The fitting quality was quantified using the root-mean-square error (RMSE), defined as
\begin{equation}
	\mathrm{RMSE}=\sqrt{\frac{1}{\mathrm{N}}\sum\limits_{k_{\parallel},n}\left[E_n^{\mathrm{extract}}(k_{\parallel})-E_n^{\mathrm{theory}}(k_{\parallel})\right]^2}
	\label{eq:6}\\
\end{equation}
where $n=\mathrm{LP},\mathrm{MP}$. The fitting window was gradually tightened during the iteration until the RMSE fell below 1 meV, ensuring a quantitative agreement between the experimental dispersion and the coupled-mode model. Then we lock in the final parameters ($g_{\mathrm{exc}}$, $g_{\mathrm{T}}$)

The Rabi splitting also cannot be expressed in an analytical form. Instead, we define it as the minimum energy separation between the interacting polariton branches at the anticrossing point:
\begin{equation}  
	2\hbar\Omega=\Delta{E_{\mathrm{anticorssing}}=\mathop{min}\limits_{k_{\parallel}}|E_i(k_{\parallel})-E_j(k_{\parallel})|}
	\label{eq:7}\\
\end{equation}
where $i,j$ denote the corresponding polariton branches involved in the coupling. The splitting values $2\hbar\Omega$ were extracted directly from the numerically calculated polariton dispersions using the fitted coupling strengths.

The Hopfield coefficients $|\mathrm{C}_{n}(k_{\parallel})|^2$, $|\mathrm{X}_{n}(k_{\parallel})|^2$, $|\mathrm{T}_{n}(k_{\parallel})|^2$ were obtained directly from the normalized eigenvectors of the Hamiltonian after numerical diagonalization, and were used to analyze the momentum-dependent evolution of the polariton compositions.

\section*{Supporting Information}

Extended data, extraction of exciton and trion PL spectra, determination of bare PhC’s photonic dispersions.(PDF)

\section*{Acknowledgements}

This work was supported by the National Key Research and Development Program of China (Grants 2024YFA1207700, and 2021YFA1400700), the National Natural Science Foundation for Young Scientists of China (Grants No.12404042). We thank Professor Mikhail M. Glazov for the fruitful discussion.

\section*{Author Declarations}
\subsection*{Conflict of Interest}
The authors have no conflicts to disclose.

\subsection*{Author Contributions}
B. L. conceived  and supervised the project. B. L. and C. G. coordinated the experimental investigations. J. Z. performed the numerical simulations,  optical measurements and fabricated the photonic crystal slabs. H. L. transferred the ML WS$_{2}$ on to PhC. B. L. and J. Z. wrote the manuscript, which all authors read and commented on. 

\section*{Data Availability}
The data that support the findings of this study are available from the corresponding author upon reasonable request.
\bibliography{manuscript}

\end{document}